\documentclass[prl,superscriptaddress,twocolumn,nopacs,letter,10.5pt]{revtex4-2}
\usepackage{graphicx,bm,times}
\usepackage{amsmath}
\usepackage{amsfonts}
\usepackage{amssymb}
\usepackage{color}
\usepackage{hyperref}
\hypersetup{
	colorlinks = true,
	allcolors = {blue}
}

\usepackage[utf8]{inputenc}
\usepackage[T1]{fontenc}
\usepackage{amsmath,amssymb}
\usepackage{lineno}
\usepackage{float}
\usepackage{siunitx}
\usepackage{soul}
\begin{document}
\title{Giant exchange splitting in the electronic structure of A-type 2D antiferromagnet CrSBr}

\author{Matthew D. Watson}
\email{matthew.watson@diamond.ac.uk}
\affiliation {Diamond Light Source Ltd, Harwell Science and Innovation Campus, Didcot, OX11 0DE, UK}

\author{Swagata Acharya}
\affiliation {National Renewable Energy Laboratory, Golden 80401 CO, USA}

\author{James E. Nunn}
\affiliation {Diamond Light Source Ltd, Harwell Science and Innovation Campus, Didcot, OX11 0DE, UK}
\affiliation {Department of Physics, University of Warwick, Coventry, CV4 7AL, UK}

\author{Laxman Nagireddy}
\affiliation {Department of Physics, University of Warwick, Coventry, CV4 7AL, UK}
\affiliation{CY Cergy Paris Université, CEA, LIDYL, 91191 Gif-sur-Yvette, France}
\affiliation{Université Paris-Saclay, CEA, LIDYL, 91191, Gif-sur-Yvette, France}

\author{Dimitar Pashov}
\affiliation{Theory and Simulation of Condensed Matter, King’s College London, The Strand, London WC2R2LS, UK}

\author{Malte Rösner}
\affiliation{Institute for Molecules and Materials, Radboud University, Heijendaalseweg 135, 6525AJ Nijmegen, The Netherlands}

\author{Mark van Schilfgaarde}
\affiliation {National Renewable Energy Laboratory, Golden 80401 CO, USA}

\author{Neil R. Wilson}
\affiliation {Department of Physics, University of Warwick, Coventry, CV4 7AL, UK}

\author{Cephise Cacho}
\affiliation {Diamond Light Source Ltd, Harwell Science and Innovation Campus, Didcot, OX11 0DE, UK}
\email{cephise.cacho@diamond.ac.uk}

\begin{abstract}
    We present the evolution of the electronic structure of CrSBr from its antiferromagnetic ground state to the paramagnetic phase above $T_N$=132 K, in both experiment and theory. Low temperature angle-resolved photoemission spectroscopy (ARPES) results are obtained using a novel method to overcome sample charging issues, revealing quasi-2D valence bands in the ground state.  The results are very well reproduced by our $\mathrm{QSG\hat{W}}$ calculations, which further identify certain bands at the X points to be exchange-split pairs of states with mainly Br and S character. By tracing band positions as a function of temperature, we show the splitting disappears above $T_N$. The energy splitting is interpreted as an effective exchange splitting in individual layers in which the Cr moments all align, within the so-called A-type antiferromagnetic arrangement. Our results lay firm foundations for the interpretation of the many other intriguing physical and optical properties of CrSBr.
\end{abstract}

\date{\today}

\maketitle


In the field of 2D magnetic semiconductors, CrSBr has been the focus of research due to its intriguing magnetic, transport and optical properties that have potential to underpin novel devices. Here we present the evolution of the electronic structure of CrSBr from its antiferromagnetic ground state to the paramagnetic phase. The ground state photoemission data, obtained using a novel method to overcome sample charging issues, is very well reproduced by $\mathrm{QSG\hat{W}}$ calculations, a self-consistent many-body perturbative approach that computes electronic eigenfunctions in presence of excitonic correlations.  The electronic structure is complex at $\mathrm{\Gamma}$, but simplifies at the X points of the Brillouin zone where the non-symmorphic lattice symmetry enforces band degeneracies. In the vicinity of these X points, in the deeper-lying valence bands one can identify pairs of states with mainly Br- and S- character. The energy splitting between them is understood as an exchange splitting within the individual layers in which the spins all align, despite the absence of a net magnetic moment in the so-called A-type antiferromagnetic arrangement. By tracing band positions as a function of temperature, we identify that this exchange splitting disappears above $T_N$, alongside a significant increase in the spectral broadening in all states. 

\begin{figure*}[t]
	\centering
	\includegraphics[width=0.94\linewidth]{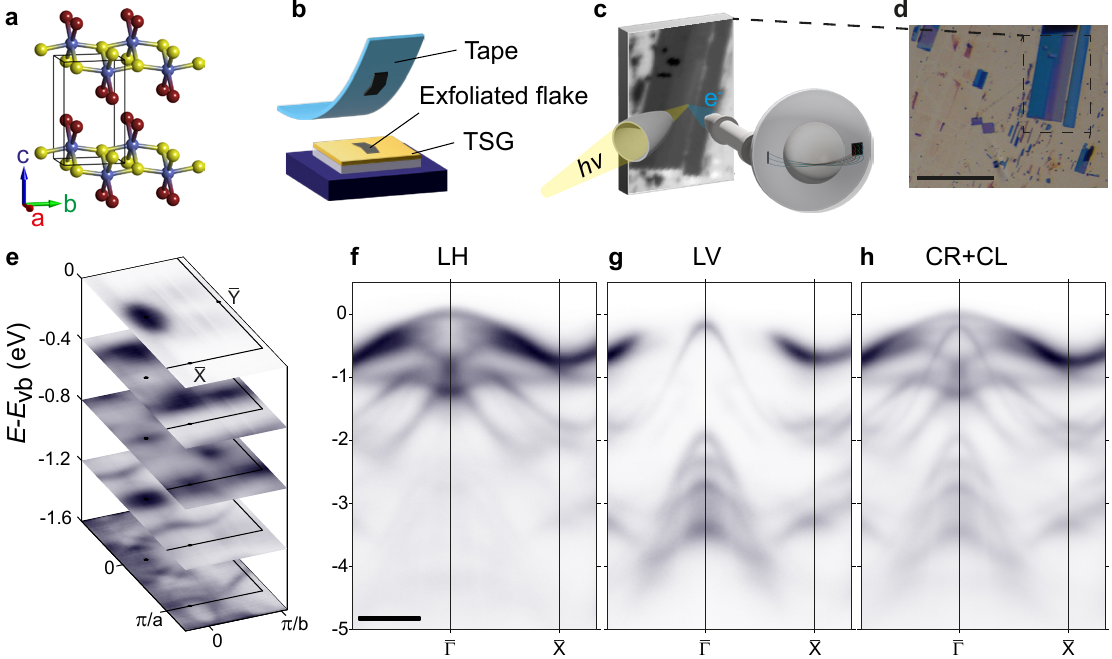}
	\caption{\textbf{Measuring the ground state electronic structure of CrSBr by micro-ARPES.} (a) Crystal structure of CrSBr (Cr = blue, S = yellow, Br = red). (b) Schematic of the last stage of sample preparation, where the tape is removed under UHV, yielding a freshly exfoliated flake on the template-stripped gold (TSG). (c) SPEM map and (d) optical microscope image of the corresponding region; scale bar is 200 $\mu$m. (e) $k_x-k_y-E$ map of the valence bands. (f-h) Dispersion along $\mathrm{\Gamma}$-X measured with linear horizontal (LH), linear vertical (LV) and the summation over two circular light polarisations (CR+CL), all at a photon energy of 53 eV and at 33~K. Scale bar is 0.5 \AA${}^{-1}$.}
	\label{fig1}
\end{figure*}

The advent of two-dimensional (2D) magnetic semiconductors, exfoliated from bulk van der Waals materials, has unveiled new opportunities to understand, combine, and tune magnetic systems for spintronics. CrSBr is a fascinating exemplar: a semiconducting A-type layered antiferromagnet that can be exfoliated down to a ferromagnetic monolayer limit \cite{lee_magnetic_2021}, it demonstrates strong exciton-phonon coupling \cite{lin_strong_2024}, exciton-coupled magnons \cite{bae_exciton-coupled_2022,diederich_tunable_2023}, exciton polaritons \cite{dirnberger_magneto-optics_2023,ruta_hyperbolic_2023}, and with indications of strong magneto-electronic coupling \cite{wilson_interlayer_2021} that leads to gate-tunable magnetoresistance \cite{telford_coupling_2022}. Working with CrSBr has several practical advantages compared with other 2D magnets: it is remarkably stable under ambient conditions \cite{ziebel_crsbr_2024}, can be readily exfoliated in large flakes whose crystal axes are easy to identify due to its crystalline anisotropy, and has a high bulk N\'{e}el temperature of $T_N$ = 132~K \cite{telford_layered_2020,goser_magnetic_1990} which is a factor of two higher than the much-studied CrI$_3$ and CrGeTe$_3$. Since the interlayer exchange interaction is much smaller than the in-plane \cite{scheie_spin_2022}, bulk ferromagnetism can be easily achieved by halide substitution \cite{telford_designing_2023}, the application of strain \cite{cenker_reversible_2022,cenker_strain-programmable_2023,diao_strain-regulated_2023}, or a modest magnetic \cite{telford_layered_2020,wilson_interlayer_2021} or electric \cite{wang_magnetic_2023} field. The strong link between spin and charge degrees of freedom, coupled to the controllable magnetic ordering, opens new spintronic opportunities that exemplify the advantages of 2D magnetic materials.

Given the convergence of interest in CrSBr, and the challenge of accurate first-principles modelling of it, there is a pressing need for experimental measurements of the electronic structure and its dependence on magnetic ordering. A previous angle-resolved photoemission spectroscopy (ARPES) study \cite{bianchi_paramagnetic_2023} characterised the electronic structure in the high-temperature paramagnetic state, however, with single-crystal samples prepared by standard routes, severe sample charging problems prohibited measurements below $\sim$200~K. At the opposite limit, ultrathin flakes of CrSBr on gold and silver crystalline substrates were measured at low temperatures, but the spectra exhibited high charge transfer causing significant population of the conduction band, and other interaction effects, such that the intrinsic low-temperature electronic structure of the insulating bulk material was not probed \cite{bianchi_charge_2023}.  Here we develop a methodology based on exfoliation of single crystals onto template-stripped gold, which enabled us to measure ‘bulk’ flakes of CrSBr at temperatures down to 33~K. We access the electronic structure in the antiferromagnetic (AFM) phase, revealing substantially more detail than the broadened high-temperature data. We use the data to test theoretical approaches and find a compelling agreement with ground state calculations using the $\mathrm{QSG\hat{W}}$ approach, where $\mathrm{\hat{W}}$ denotes the screened Coulomb interaction computed including excitonic vertex corrections (ladder diagrams) by solving a Bethe–Salpeter equation (BSE). The lower-lying valence bands with Br and S character exhibit a giant exchange splitting in the AFM phase. Focusing on a particular pair of such states around the X point, we find that their energy splitting of 250 meV in the ground states disappears above $T_N$, further confirming their assignment as exchange-split pairs of bands. The combination of experimental electronic structure determination and state of the art theoretical calculations establish a firm framework for the understanding, and eventual utilization, of the intriguing properties of CrSBr.  

\textbf{Results} \\
Our sample preparation route is a variation on the theme of using clean and flat Au surfaces for the exfoliation of 2D materials \cite{huang_universal_2020,bianchi_charge_2023,grubisic-cabo_situ_2023}. We make use of the long-established technique of producing ``template-stripped" Au \cite{hegner_ultralarge_1993,vogel_as_2012}, using both silicon and mica as a template. In an argon glove box, we exfoliate CrSBr onto the freshly-exposed template-stripped Au surface leaving the transfer tape attached, move it to the ultra-high vacuum (UHV) system minimizing exposure to air, and finally remove the transfer tape in the load lock of the UHV system immediately before measurement, as shown schematically in Fig.~\ref{fig1} and further detailed in the supplementary information (SI). The benefit of this approach is that several samples can be prepared in advance, with no \textit{in-situ} preparation required. We propose that this technique should be suitable for many 2D materials, and specifically opens a practical route to measuring ARPES on 2D magnetic semiconductors below their ordering temperatures, as we do here. 

\begin{figure}[t]
	\centering
	\includegraphics[width=\linewidth]{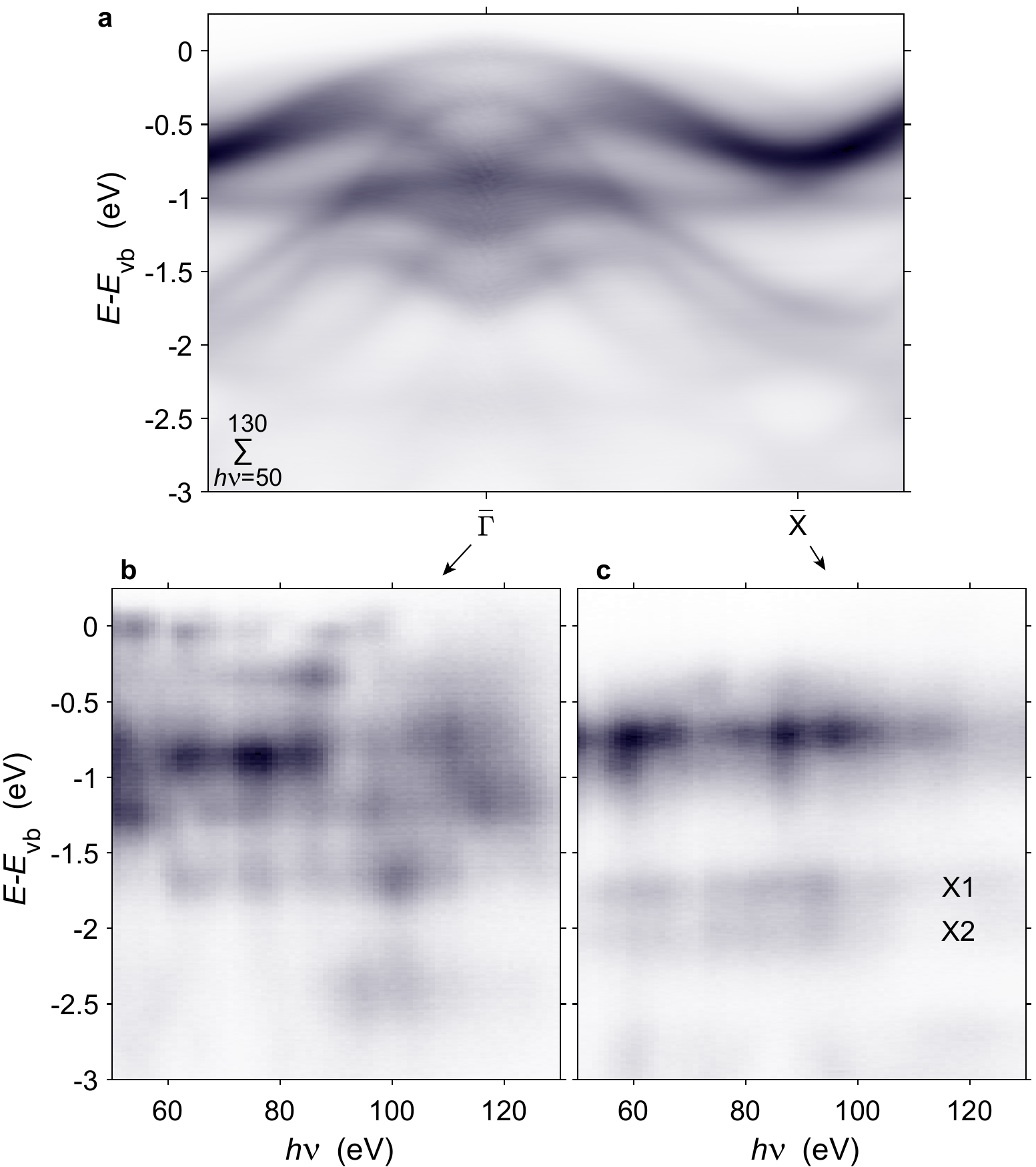}
	\caption{\textbf{Photon energy-dependent ARPES.} (a) Summation of ARPES spectra measured with $h\nu$=50 to 130 eV (step of 1 eV, LH polarisation) at 33~K. (b,c) photon energy dependence of the energy distribution curves (EDC) at $\mathrm{\Gamma}$ and X respectively. Note that the data is referenced to photon energy dependent measurements of the Fermi edge on the template-stripped gold substrate, with $E_{vb}=E_F-1.73$ eV.}
	\label{fig2}
\end{figure}

\begin{figure*}[t]
	\centering
	\includegraphics[width=0.75\linewidth]{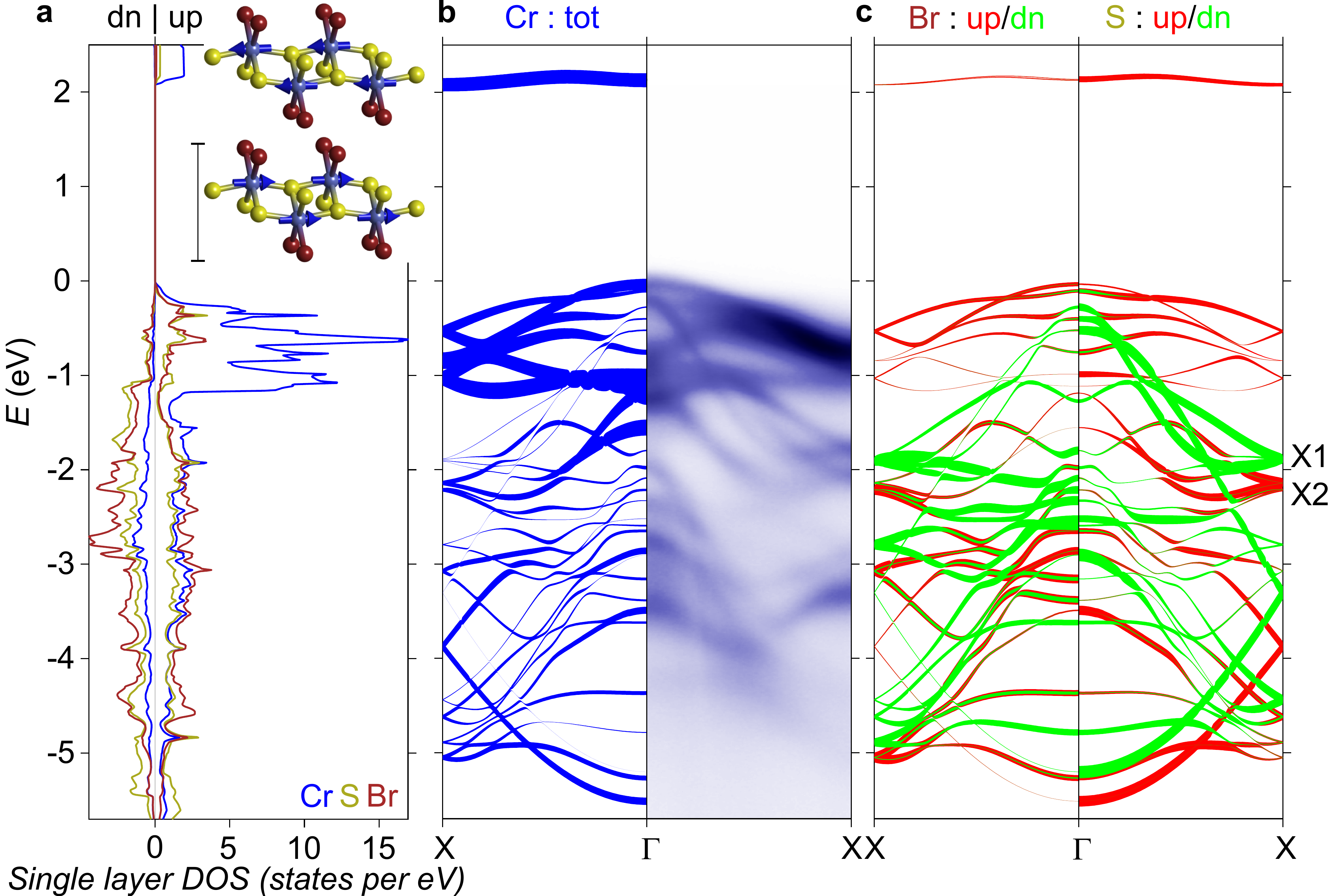}
	\caption{$\mathbf{QSG\hat{W}}$\textbf{ calculations.} (a) Atom-resolved (Mulliken projection) partial density of states from $\mathrm{QSG\hat{W}}$ calculations. The projection here, as well as panel (c), runs over the atoms within a single layer (inset) rather than the whole AFM unit cell. (b) Comparison of the calculated band dispersions along $\mathrm{\Gamma}$-X, where the linewidth represents the total Cr weight (i.e. summing over the spin index), with ARPES data (measured with $h\nu{}$=53~eV at 33~K, same data as Fig.~1(h)). (c) Calculated dispersions plotted with spin-resolved (red=up,green=down) weight of the Br (left side) and S (right side) character. Two pairs of bands with dominantly S and some Br character at the X point are highlighted and labelled as X1 and X2.}
	\label{fig3}
\end{figure*}

CrSBr has an orthorhombic structure as shown in Fig.~\ref{fig1}(a), in the space group \textit{Pmnm} (no. 59) \cite{lopez-paz_dynamic_2022}. Both single crystals and exfoliated flakes typically have ribbon-like morphology with large aspect ratios, with the longer dimension along the crystallographic \textit{a} axis \cite{cenker_reversible_2022}. The preparation yields several such flakes, but without detailed prior knowledge of the sample morphology, we started by performing scanning photoemission mapping (SPEM) making use of the capillary mirror at the the I05 beamline, Diamond Light Source, to achieve a beam spot of $\sim$4 \textmu m, as shown schematically in Fig.~\ref{fig1}(c,d). Good candidate regions were identified by the appearance of sharp valence bands without charging effects, where typically the thickness of the sample is $\sim$10-30 nm  (Supplementary Figures 1,2). In this regime, the sample is thick enough (20 nm $\approxeq$ 25 unit cells) that the magnetic properties should be essentially bulk-like (given that the $T_N$ converges to the bulk value with just 6 layers \cite{ziebel_crsbr_2024}), while electronically there are enough layers that quantisation of the $k_z$ axis is negligible, and furthermore the surface-sensitivity of the photoemission ensures there is no signal from the gold substrate.   

The dispersion measured by ARPES in the high-symmetry $\rm \Gamma$-X direction shown in Fig.~\ref{fig1}(f-h) is modulated by strong matrix element effects: in linear horizontal (LH) polarisation, the spectrum is dominated by a manifold bands of bright states dispersing within $\sim$1~eV of the valence band maximum, while below this we find further sharp valence bands, some of which are more visible in the measurement in linear vertical (LV) polarisation. The summation over the two circular light polarisations in Fig.~\ref{fig1}(h) works well in this case to give a balance of spectral weight on many bands, though variation of photon energy is also required to observe some states not observed here. An important observation is that the band structure substantially simplifies at the X point, where the bands appear to bunch together. 

The effective dimensionality of CrSBr has been a topic of some discussion, since the optical responses are very strongly anisotropic \cite{wilson_interlayer_2021,ruta_hyperbolic_2023} and some reports have emphasised one-dimensional perspectives \cite{wu_quasi-1d_2022,klein_bulk_2023}. While the conduction band dispersions are indeed calculated to be extremely anisotropic and could therefore yield an apparent one-dimensionality in experimental methods which are sensitive to the conduction states, with ARPES we detect only the occupied valence bands. We find that most bands, including the uppermost valence bands, disperse substantially in both the X and Y directions in Fig.~\ref{fig1}(e). Varying the photon energy probes the dispersion in the out-of-plane $k_z$ direction. Our measurements in Fig.~\ref{fig2} show 
the band positions are unchanged with photon energy. Hence, the sum over spectra obtained with varying photon energy in Fig.~\ref{fig2}(a) shows sharp features, reflecting the 2D nature of at least most of the bands. The absence of any clear periodic features precludes a confident mapping from $hv$ to $k_z$ in the nearly free electron final state approximation, but further confirms a quasi-2D character of the electronic structure in the ground state, in line with calculations for the AFM phase \cite{bianchi_paramagnetic_2023}. Combined, these results give an overall picture of a complex multi-band, but quasi-2D electronic structure of the valence bands.

Given that sample charging usually prohibits low-temperatures ARPES measurements of 2D magnetic semiconductors, the high quality data obtained with our novel approach yields a rare opportunity to rigorously test theoretical approaches to calculating the ground state electronic structure. In Supplementary Figure 4 we overlay the experimental $\mathrm{\Gamma}$-X dispersion with various \textit{ab-initio} methods: a standard density functional theory (generalised gradient approximation, GGA) calculation \cite{blaha_wien2k_2020}, a calculation using the mBJ potential \cite{koller_improving_2012}, $\mathrm{QSGW}$, and $\mathrm{QSG\hat{W}}$, where the last differs from $\mathrm{QSGW}$ by the addition of ladder diagrams to \textit{W} via a  two-particle Bethe-Salpeter equation \cite{cunningham_2023}. We find that ``vanilla" GGA yields an unphysically small band gap, and the dispersions calculated using the mBJ potential match poorly, but the QSGW calculation gives a reasonable approximation of the measured valence bands. However, our ground state data are of sufficiently high resolution to demonstrate that the agreement with experiment is further improved in the $\mathrm{QSG\hat{W}}$ approach (see detailed discussion in SI). Indeed in Fig.~\ref{fig3}(b), we find that not only are the band dispersions well-matched between the experiment and $\mathrm{QSG\hat{W}}$ calculations, but also the projection of Cr weight onto the bands captures many features in the distribution of spectral weight in the experimental data taken with $h\nu{}$=53 eV, where the cross section for photoemission from Cr dominates over S or Br.

A robust finding of the $\mathrm{QSG\hat{W}}$ calculations in Fig.~\ref{fig3} is the prediction of a single-particle band gap of 2.09 eV. The experimental perspective on 
the band gap, remains a topic of substantial debate in the literature, however. Transport measurements indicated an activation gap of only $\sim$93 meV \cite{telford_layered_2020}- presumably due to the pinning of the chemical potential by donor states. Meanwhile tunneling experiments reported a gap of 1.5(2) eV \cite{telford_layered_2020,klein_bulk_2023}. Several papers have focused on excitons with very sharp linewidths, with the most prominent one observed by absorption and photoluminescence at 1.37 eV \cite{wilson_interlayer_2021,klein_bulk_2023,ruta_hyperbolic_2023}, while optical absorption and photoacoustic spectroscopy seem to coalesce around a band gap of $\sim$1.2 eV at room temperature \cite{linhart_optical_2023}. On the other hand, the previous ARPES report \cite{bianchi_paramagnetic_2023} suggested the gap might be closer to 2 eV. Our experimental results align closely with this: the measured valence band maximum (peak of EDC) is 1.6-1.8 eV below the Fermi level (varying slight with sample, position, and temperature) as measured on adjacent regions of the template-stripped gold. That there is no evidence of any population of the conduction band even in our measurements at 200~K implies that the fundamental gap is $\gtrsim$ 0.1 eV larger than 1.8 eV. These disparate results can be potentially reconciled by considering that the fundamental gap is indeed $\sim$ 2 eV - aligning with our calculations - but that there are excitonic states with large binding energies $\sim$ 0.6 eV \cite{ruta_hyperbolic_2023,shao2024exciton,smolenski_large_2024} that dominate the optical response at energies below the fundamental single-particle band gap, analogous to the scenario proposed for CrX$_3$ (Supplementary materials of \cite{grzeszczyk2023strongly}). 

With the excellent overall agreement between the theory and experiment established, we seek to build a global picture for the ground state electronic structure of CrSBr. First, we identify the manifold of relatively flat states within 1.25~eV of the valence band maximum, which are also the brightest bands seen experimentally, as bands deriving from the occupied Cr $t_{2g}$ spin-majority orbitals, the basis of the S=$\frac{3}{2}$ magnetic moments. The calculated DOS in Fig.~\ref{fig3}(a) confirms that these are fully spin-polarised within a single layer.

The lower bands have primarily S and Br character but, interestingly, the (single-layer) spin-projected calculations in Fig.~\ref{fig3}(c) identify bands that manifest as pairs with near-parallel dispersions, where the spin-up branches (aligned to the Cr $t_{2g}$ bands) are at lower energies. While the band dispersions at $\mathrm{\Gamma}$ are somewhat complex to interpret, the situation simplifies significantly at the X point. We highlight two particular states marked as X1 and X2: these have mostly S but also some Br character, and a non-zero contribution of Cr weight on X2 (the spin-up branch). The bands which disperse towards X1 and X2 are very similar, and can therefore be understood as spin-split pairs. The calculations point to several such pairs, but the X1-X2 pair is the one at the highest energy, is most easily identifiable experimentally, and is already shown to be $k_z$-independent in Fig.~\ref{fig2}(c).

The spin-split bands should be more precisely described as an exchange-splitting, a phenomenon normally associated with \textit{ferromagnetic} order. A classic example is EuO, considered a prototypical Heisenberg ferromagnet: when the Eu$^{2+}$ $4f$ moments order ferromagnetically below $T_C$=69~K, calculations predict and experiments have confirmed that the occupied valence bands with primarily O $2p$ character split into spin-up and spin-down branches \cite{miyazaki_direct_2009,heider_temperature-dependent_2022}. The observed energy difference of $\sim$0.2 eV is far too large to be described as a Zeeman splitting, but is rather understood as a consequence of magnetic exchange due to finite hybridisation with the Eu $4f$ as well as the nominally unoccupied $5d$ and $6s$ conduction states. 

In comparison to EuO, CrSBr has a larger unit cell, more numerous bands, lower symmetry including non-symmorphic elements, and strong spin-orbit coupling at least for states with Br character, all of which make the situation more complex, but the biggest difference is that CrSBr is antiferromagnetic. Nevertheless, since the interlayer coupling is weak and monolayer CrSBr is ferromagnetic, conceptually there can still be effective exchange splittings within each layer. The spin-up branch of ligand states can hybridise with Cr states within the layer - and therefore delocalise and lower their kinetic energy - but the counterpart spin-down Cr states are at much higher energy (or located in the next layer), suppressing any possible hybridisation. As a result, the ligand bands split into two, with the spin-up branch having lower energy. The magnitude of the exchange splitting will vary dependent on the geometry and wavefunction of the relevant orbitals, but this pattern is repeated across all valence bands in the vicinity of the X points in the calculations in Fig.~\ref{fig3}(b,c): bands come in pairs, where the spin-up branch is at lower energy, and has some admixture of Cr weight. 

\begin{figure*}
	\centering
	\includegraphics[width=0.75\linewidth]{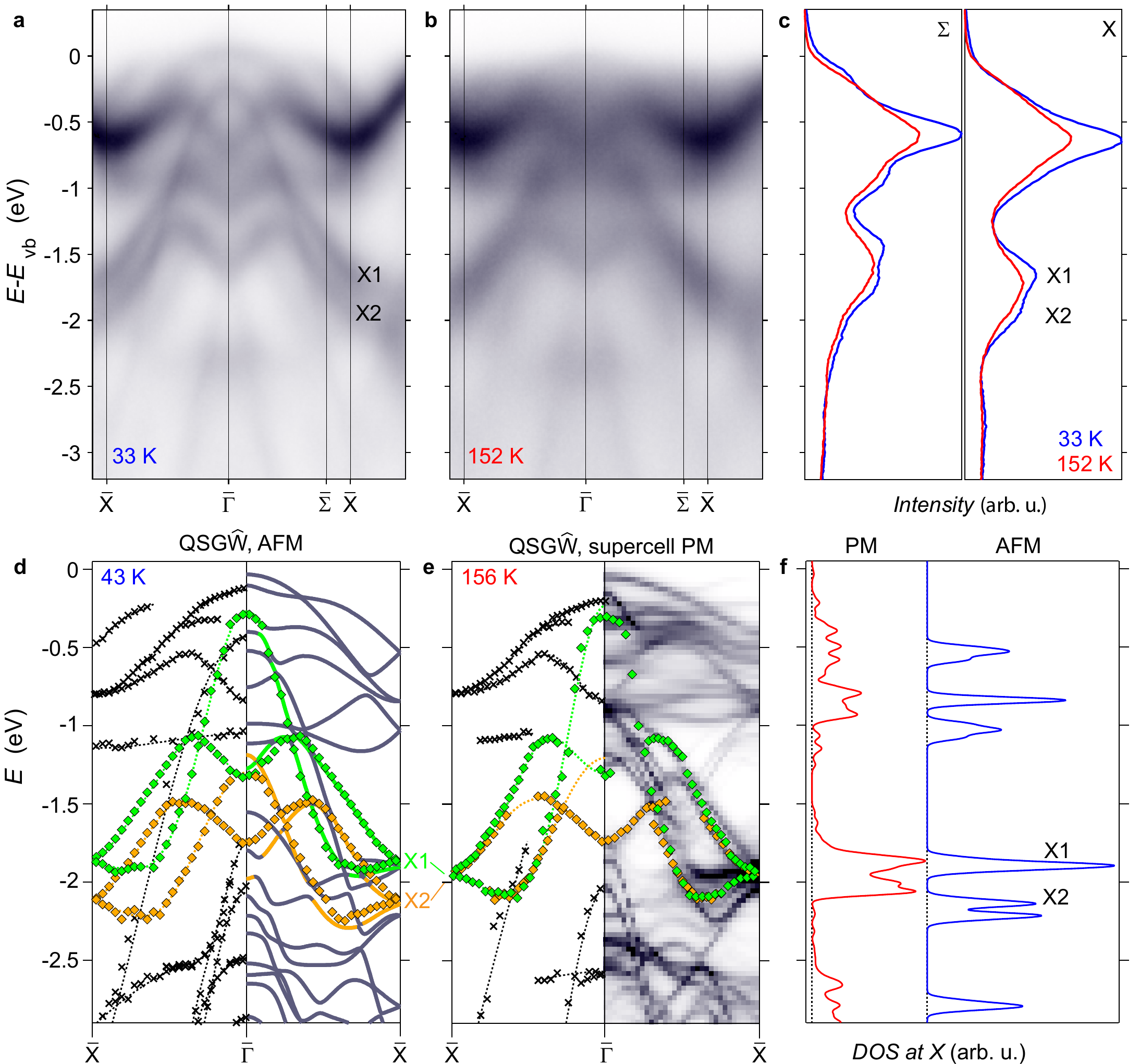}
	\caption{\textbf{Collapse of the exchange splitting in the paramagnetic state.} (a) ARPES measurements with hv=90 eV and LH polarisation, at 33 K and (b) at 152 K, above $T_N$ = 132 K. (c) EDCs at the $\Sigma$ ($k_x$ = 0.72 \AA$^{-1}$) and X points, both showing a double-peak structure that is the energy scale of the exchange splitting, which collapses at high temperature. (d) Dispersions as obtained from band tracing, combining results from 53 eV and 90 eV data in both linear light polarisations. Note that bands observed beyond the X points are backfolded into the first Brillouin zone, and dotted lines are guides to the eye. The pairs of bands that disperse towards X1/X2 and are exchange-split in the AFM phase are highlighted in green and orange. The data is overlaid on the $\mathrm{QSG\hat{W}}$ bands (as Fig.~3). Note that here the energy scale is set to 0 for the top of the calculated valence band, and the energies of the band tracing are shifted such that the top of the green band aligns (avoiding ambiguity in the structure of the measured valence band maximum). (e) Band tracing in the paramagnetic phase, using the energy axis defined at low temperature, and overlaid on $\mathrm{QSG\hat{W}}$ 2$\times{}$2$\times{}$2 supercell calculations with randomised spin moments on the Cr sites, as described in main text. (f) Comparison of the partial DOS at the X points in the $\mathrm{QSG\hat{W}}$ AFM ground state and disordered supercell calculations, also showing the collapse of the separate peaks labelled X1 and X2 in the disordered phase.}
	\label{fig4}
\end{figure*}

An intriguing question is how the manifestation of exchange splitting, which we have established in the ground state, evolves with temperature. Since these states have primarily S and some Br character, however, the experimental conditions for optimal observation differs somewhat from the Cr states. While 53 eV, as used in Fig.~\ref{fig1} and \ref{fig3}, highlights primarily the Cr weight, at 90 eV, as used in Fig.~\ref{fig3}(a), although the Cr $t_{2g}$ states are still brighter, we find relatively more intensity on the bands at X1/X2 (see also Fig.~\ref{fig2}(c)) and observe a clear splitting of bands at X1/X2 in the data in Fig.~\ref{fig4}(a). In the energy distribution curves (EDCs) taken at both X and $\Sigma$ ($k_x$=0.72 \AA$^{-1}$) and shown in Fig.~\ref{fig4}(c), the 33~K data show a two-peak structure with a separation of $\approx$250 meV. In contrast, above $T_N$=132~K at 152~K, there is a single-peak structure. This constitutes clear evidence that the exchange splitting collapses - or at least is drastically suppressed - above $T_N$.

The ARPES data in Fig.~\ref{fig4}(a,b) is asymmetric around X, and the intensity follows different dispersions in the $\Gamma_{00}$-X and X-$\Gamma_{10}$ sectors, despite these being equivalent in terms of the calculations. We understand this to be an effect of the glide-mirror symmetry, as described further in Supplementary Figure 3. This effect, together with the effect of light polarisation and photon energy variation, means that not all bands are simultaneously observed in any given experimental condition. Thus, to account for as many dispersions as possible, in Fig.~\ref{fig4}(d,e) we present the results of band-tracing analysis, combining band positions extracted from curvature analysis of four data sets at each temperature (53 eV and 90 eV photon energy, and both LH and LV polarisations; for more detail on the band tracing see Methods and Supplementary Figure 5). With this analysis, we can make a 1-1 assignment of many (but not all) of the states seen at low temperature with bands predicted by the calculations, as shown in Fig.~\ref{fig4}(d). We identify that each of the X1 and X2 peaks in the EDC at the X point corresponds to a pair of dispersions that become degenerate at X, and we highlight these dispersions with green and orange markers respectively. The experimental dispersions follow segments of the calculated bands very closely (solid green/orange lines in Fig. Fig.~\ref{fig4}(d)). At high temperature, as shown in Fig.~\ref{fig4}(e), the orange and green bands merge, at least in the vicinity of $\mathrm{\bar{X}}$. Since these bands are already identified as spin-split pairs in the ground state (see Fig 3c), this further signifies the collapse of the exchange splitting in the paramagnetic phase.

As well as the collapse of the exchange splitting, the overall broadening of the spectra increases strongly between the measurements above and below $T_N$, although band-like features are retained even above $T_N$. 
In the paramagnetic phase, the system continuously samples different magnetic configurations and associated band structures. Hence, the measured spectral function effectively averages over these and broadened spectra are to be expected. We do not fully account for such dynamical spin fluctuations within our $\mathrm{QSG\hat{W}}$ framework, but we can consider static magnetic disorder to evaluate the magnitude of this effect. We thus performed calculations in a 2$\times{}$2$\times{}$2 supercell (of a 6-atom unit cell, or 2$\times{}$2$\times{}$1 supercell of a 12-atom AFM unit cell) with randomised orientation of the spin moments on the Cr sites. This creates an four-fold multiplication of the number of bands in the calculation. So, for comparison with the experimental results in Fig.~\ref{fig4}(e), we plot the supercell band structure with spectral weight projected in the first Brillouin zone. These calculations give a qualitative indication of how magnetic disorder can cause the broadening of spectral weight - and not just on the Cr-derived states. However, certain band-like structures remain, including the build-up of spectral weight along dispersions that come close to the peak intensities in the experimental data (collapsed green and orange data points in Fig.~\ref{fig4}(e)). Further, upon plotting a partial density of states (DOS) at the X point of the magnetically disordered phase in Fig.~\ref{fig4}(f), we find that calculation mimics the experimental result, in that the merging of the peaks X1 and X2 is one of the most notable changes between the AFM and PM phases. The collapse of exchange splitting is therefore also captured in the magnetically disordered calculation.

\textbf{Discussion}
These results shed light on the intricate interplay between magnetic ordering and electronic structure in CrSBr that underpins many of the previously observed phenomena. The low-temperature spectra, whose acquisition was made possible through a simple development in sample fabrication, demonstrate split pairs of bands which, through comparison with theory, are assigned as due to exchange splitting. Importantly, this exchange splitting disappears at higher temperatures where the magnetic disorder also induces broadening of the spectra.  The strong magneto-electronic coupling in CrSBr should be probed further by measurements - including possible spin-resolved ARPES studies - in the ferromagnetic state, accessible through strain \cite{cenker_reversible_2022,cenker_strain-programmable_2023,diao_strain-regulated_2023} or chemical substitution \cite{telford_designing_2023}.

Although the conduction band edge is not observed, both $\mathrm{QSG\hat{W}}$  calculations and the experimental spectra support the attribution of a band gap of $\sim$2 eV,  a subject of debate in the literature. A further argument in favour of a $\sim$2 eV gap comes from the optical absorption: with light polarised along the $b$-axis, a sharp feature in optical absorption is observed at 1.37 eV \cite{wu_quasi-1d_2022,ruta_hyperbolic_2023}, seen also prominently in photoluminescence \cite{wilson_interlayer_2021,klein_bulk_2023,wang_magnetically-dressed_2023}, but with light polarised along the $a$-axis the first bright optical absorption peak observed experimentally is at $\sim$2 eV \cite{wu_quasi-1d_2022}, which we understand as an allowed optical transition at the band-edge \cite{shao2024exciton}, and thus representative of the fundamental gap rather than bound states. We suggest that time-resolved ARPES or inverse photoelectron spectroscopy could be highly informative in pursuit of the fundamental gap in CrSBr. 

From an \textit{ab initio} perspective, our results confirm the deficiency of DFT-based approaches in 2D magnetic semiconductors. While $\mathrm{QSGW}$ gives a reasonable match to the experimental dispersions (SI), the inclusion of BSE ladder diagrams self-consistently in $\mathrm{QSG\hat{W}}$ leads to improved and excellent overall agreement with the valence band measurements shown in Fig.~\ref{fig3}. As well as this, the approach also naturally and accurately yields the two-particle bound states highly relevant for optical transitions in CrSBr \cite{shao2024exciton}. These $\mathrm{QSG\hat{W}}$ calculations should thus be considered the gold standard for \textit{ab initio} calculations on 2D magnets, and particularly those based on Cr$^{3+}$ \cite{acharya_theory_2023,acharya2022real,acharya2021electronic}. 

The energy scale of the exchange splitting we observe, $\approx$ 250 meV, at first glance seems to be large in comparison with the in-plane magnetic exchange Hamiltonian parameters $J_{1,2,3}$ which were found to be on the order of 1-3 meV from fits to the magnon dispersions \cite{scheie_spin_2022} and calculations \cite{rudenko_dielectric_2023}. If these energy scales both reveal something about magnetic exchange, the disparity of their magnitude needs explanation. The first consideration is that any energy gain from the exchange splitting needs to be be calculated as a sum over all bands: the majority spin band shifts to lower energy in the ordered phase, but if the minority spin moves up in energy - as appears to be the case from our analysis in Fig.~\ref{fig3}(d,e) - the net energy gain would be considerably less. Further, the energy gain needs to be considered as a sum over all momenta; here we have focused on the X point since this is where we have clear experimental signatures, but all states are relevant if considering changes in the total energy. Moreover, the sum of the various exchange Hamiltonian parameters, considering also the multiplicity of the nearest neighbours, is at least an order of magnitude larger than the individual $J_{ij}$ values. Thus, while we do not direct association between the spectral signatures of exchange splitting and any specific parameter in the spin exchange Hamiltonian, they are in a broader sense two sides of the same coin. 

More generally, following the arguments above, in ARPES measurements of insulating ferromagnets (or A-type antiferromagnets as in CrSBr), it is natural to expect to observe shifts and splittings of certain bands that may be on significantly larger energy scales (e.g. $\approx$250 meV at X1/X2) than would be naively suggested by the temperature of the phase transition ($k_BT_N$=11.4 meV). ARPES measurements can thus give an insight into the raw energy gain that underpins most forms of magnetic superexchange in semiconductors \cite{kanamori_superexchange_1959}. However, with the exception of CrGeTe$_3$ \cite{watson_direct_2020} there have been surprisingly few such ARPES studies attempted - in part, due to the experimental challenge of sample charging at low temperatures. Our methodology therefore opens the door to many future important insights into magnetic exchange in quasi-2D systems by ARPES. 
\\
\\
\textbf{Methods}\\
\textbf{Experimental Methods.}
CrSBr crystals were obtained commercially from HQ Graphene and exfoliated onto template stripped gold under high vacuum conditions (see Supplementary Figure 1 for an illustration and further detail). Micro-ARPES measurements were performed at the I05 beamline \cite{hoesch_facility_2017} with a base temperature of 33~K. We then typically aligned samples such that the $a$ axis - or $\rm \Gamma$-X in $k$-space - is parallel to the entrance slit of the analyser, while the perpendicular direction $\rm \Gamma$-Y is accessed by rotating the analyser around the sample position. The energy scale of the data presented is referenced to the $E_{vb}$, i.e. the first peak in the EDC through $\rm \bar{\Gamma}$ at low temperatures. For the high temperature data in Fig.~\ref{fig4}(b), the $E_{vb}$ is not changed from the low temperature one. For example spectra referenced to $E_F$, the Fermi level of the gold substrate, see Supplementary Figures 2 and 5. The band tracing procedure using in Fig.~\ref{fig4} is based on curvature analysis and comparing and combining different photon energies and light polarisations, as further detailed in Supplementary Figure 5. The SI also contains a discussion of the impact of glide-mirror symmetry on the measurements. 


\textbf{Theory Methods.}
Theory: The Quasiparticle Self-Consistent GW approximation is a self-consistent form of Hedin's GW approximation~\cite{hedin65}.  Self-consistency eliminates the starting point dependence, and as a result the discrepancies are much more systematic than conventional forms of GW~\cite{mark06qsgw,pashov2020questaal}.  The great majority of such discrepancies in insulators originate from the omission of electron-hole interactions in the RPA polarizability.  By adding ladders to the RPA, electron-hole effects are taken into account.  Generating \textit{W} with ladder diagrams has several consequences; most importantly, perhaps, screening is enhanced and \textit{W} reduced.  This in turn reduces fundamental bandgaps and also valence bandwidths.  Agreement with experiment in both one-particle and two-particle properties is greatly improved. The theory and its application to a large number of both weakly and strongly correlated insulators is given in Ref.~\cite{cunningham_2023}. The importance of self-consistency in both $\mathrm{QSGW}$ and $\mathrm{QSG\hat{W}}$ for different materials have been explored \cite{acharya2021importance}.

For bulk CrSBr in the AFM phase with 12-atom unit cell, we use a=3.504 \AA, b=4.738 \AA. Individual layers contain ferromagnetically polarized spins pointing either along the +b or -b axis, while the interlayer coupling is antiferromagnetic. The single particle calculations (LDA, and energy band calculations with the static quasiparticlized $\mathrm{QSGW}$ and $\mathrm{QSG\hat{W}}$ $\Sigma$(k)) are performed on a 10$\times$7$\times$4 k-mesh while the relatively smooth dynamical self-energy $\Sigma(\omega)$ is constructed using a 6$\times$4$\times$2 k-mesh.  The $\mathrm{QSGW}$ and $\mathrm{QSG\hat{W}}$ cycles are iterated until the RMS change in the static part of quasiparticlized self-energy $\Sigma(0)$  reaches 10$^{-5}$ Ry. The two-particle Hamiltonian that is solved self-consistently to compute both the $\Sigma$ and the excitonic eigenvalues and eigenfunctions, contained 26 valence bands and 9 conduction bands.

For the bulk PM phase, 48 atoms were included in the supercell and 48 interstitial sites were also added to augment the basis with floating orbitals. For the PM phase, local spin orientations were arranged in a quasi-random configuration to minimize the difference between the quasirandom and true random site correlation functions. An objective function composed from 480 pair and 384 triplet functions was minimized, following the approach by Zunger et al.\cite{zunger}  Four completely independent quasirandom configurations were made and $\mathrm{QSG\hat{W}}$ calculations were performed independently for all cases. The DFT total energies for each of these quasirandom configurations were different.  For the PM configurations, energy band calculations with the static quasiparticlized $\mathrm{QSGW}$ and $\mathrm{QSG\hat{W}}$ $\Sigma$(k)) are performed on a 9$\times$9$\times$6 k-mesh while the relatively smooth dynamical self-energy $\Sigma$ is constructed using a 3$\times$3$\times$2 k-mesh. In consistency, with the 12-atom AFM case, for the 48-atom PM supercell, 104 valence and 36 conduction bands were included in the two-particle Hamiltonian~\cite{ruta_hyperbolic_2023,bianchi_paramagnetic_2023}. For further rigorous comparisons between AFM and PM phase, the 12-atom AFM unit cell was raised to the level of a 2$\times$2$\times$1 supercell containing 48 atoms and calculations were performed using the exact same parameters as that of the 48-atom PM supercell. The necessity and sufficiency of such theories in describing both one- and two-particle transitions in a large class of antiferromagnets in their ordered and disordered phases have been described in more details in prior works~\cite{acharya_theory_2023,acharya2022real,acharya2021electronic}.

Brillouin zone folding of supercells: we developed a method to resolve an eigenstate $\psi^{n\mathbf{k}}$ of the supercell into linear combinations of Bloch functions $\phi^{i\mathbf{k}_1}\dots\phi^{i\mathbf{k}_N}$ of the primitive cell.   Related approaches have been implemented previously in a model context~\cite{Dargam97,Medeiros14}, and a density-functional framework~\cite{Wang20}. Taking a supercell as a perturbation to a reference constructed from an \textit{N}-fold replication of a primitive unit cell, the Brillouin zone of the primitive cell is folded into the supercell so that \textit{N}  \textbf{k}-points in it (points $\mathbf{k}_1\dots\mathbf{k}_N$), become reciprocal lattice vectors $\mathbf{G}_1\dots\mathbf{G}_N$  of the supercell.  This decomposition, which can be cast as a \textbf{k}- and energy-dependent self-energy of the primitive cell, provides an alternative scheme to obtain nonlocal self-energies emerging from spin or lattice disorder.\\

\clearpage

\textbf{Data Availability} \\
The data that support the findings of this study are available from the corresponding author upon reasonable request. \\

\textbf{Acknowledgments}\\
We thank T. K. Kim, A. Louat, P. Hofmann, and D. Cobden for insightful discussions. We acknowledge Diamond Light Source for time on beamline I05 under proposals SI33317 and SI34489. We acknowledge support from the EPSRC through grant EP/T027207/1. L.N-R was supported by a EUTOPIA PhD Co-tutelle Programme.
S.A., D.P. and M.v.S were supported the by the Computational Chemical Sciences program within the Office of Basic Energy Sciences, U.S. Department of Energy under Contract No. DE-AC36-08GO28308. S.A., D.P. and M.v.S acknowledge the use of the National Energy Research Scientific Computing
Center, under Contract No. DE-AC02-05CH11231 using NERSC award BES-ERCAP0021783 and the computational resources sponsored by the Department of Energy's Office of Energy
Efficiency and Renewable Energy and located at the National Renewable Energy Laboratory.
\\
\\
\textbf{Author Contributions}\\
J.N. led the preparation and characterisation of the samples. L.N-R. and J.N. prepared the template-stripped gold. J.N., L.N-R., M.W., N.W. and C.C. performed the micro-ARPES experiments, and M.W., J.N. and C.C. analysed the data. S.A. performed the QSGW and BSE calculations, in collaboration with D.P., M.R. and M.v.S. The paper was written by M.W., S.A., N.W. and C.C. with input from all coauthors. N.W. and C.C. led the overall direction of the experiments.
\\
\\

\textbf{Competing Interests Statement}\\
The authors declare no competing interests.

\clearpage

\textbf{References} \\


\end{document}